\documentclass[12pt,]{article}
\usepackage{amsmath}
\usepackage{amsfonts} \usepackage{amssymb}

\numberwithin{equation}{section}

\begin{document}      

\footnotesize

\title{
Peeling or not peeling 
$-$ 
is that the question ?\footnote{Extended version of a talk given at the symposium honoring Piotr T. Chru\'sciel on the occasion of his 60th birthday, 17th - 18th August 2017.}}

\author {Helmut Friedrich\\
Max-Planck-Institut f\"ur Gravitationsphysik\\
Am M\"uhlenberg 1\\ 14476 Golm, Germany }

\maketitle                
 
\begin{abstract}

The concepts of isolated self-gravitating system, asymptotic flatness and asymptotic simplicity are reconsidered, 
various related results are discussed and put into perspective,  basic open questions
are discussed.

\end{abstract}

\noindent
MSC  83C05, 83C30, 83C35



\section{Introduction}

\vspace{.3cm}

The direct measurement of gravitational radiation \cite{abbott-et-al-2016} must be seen  a triumph for  experimental  as well as for theoretical physics.
In view of the graphs showing the impressive coincidence of the measured and the calculated radiation signals, one might think that everything comes to a conclusion and is understood now. 
But new questions will come up (see \cite{creswell-et-al:2017} for an example) and may require more precise statements. It will thus still be worthwhile to reconsider questions that have been left open in the  theory of gravitational radiation. After giving an outline of the basic ideas and results concerning the underlying mathematical structure I shall discuss some of the remaining unresolved problems.

\vspace{.3cm}

\noindent
Following a gestation period of several years, involving many workers,  the basic setting for the analysis of gravitational waves was  proposed in the early 1960's by H. Bondi et al. \cite{bondi:et.al}, R. Sachs  \cite{sachs:1962}, E.T. Newman and R. Penrose \cite{newman:penrose}. It requires: The idealization of an {\it isolated self-gravitating system}, the analysis of solutions to Einstein's field equations which are {\it asymptotically flat in null directions}, the control of the evolution by Einstein's field equations on large scales, the control of the geometry on large scales, precise asymptotics at space-like and null infinity, and the definition of  physical concepts related to physical observations  `far away from the system'.

\vspace{.3cm}

\noindent
Carving out  the role of  null and conformal geometry in the analysis of  space-time structures in the 
large, R. Penrose \cite{penrose:1963} combined the various ideas in the elegant geometric concept of {\it asymptotic simplicity}, which characterizes the expected asymptotic behavior by the requirement that {\it the conformal structure be smoothly extendable across null infinity}.

\vspace{.3cm}

\noindent
The basic model is provided by Minkowski space $\hat{M} = \mathbb{R}^4$, $\hat{g} = -dt^2 + dr^2 + r^2\,d\sigma^2$, given here in spatial polar coordinates with $d\sigma^2$ denoting the standard line element on $\mathbb{S}^2$ and coordinates $t \in \mathbb{R}$ and $r \ge 0$.
Performing the coordinate transformation
\[
t(\tau, \chi) = \frac{\sin \tau}{\cos \tau + \cos \chi}, \quad 
r(\tau, \chi)  = \frac{\sin \chi}{\cos \tau + \cos \chi}, 
\]
and rescaling with the conformal factor
$ \Omega =  \cos \tau + \cos \chi = \frac{2}{\sqrt{1 + (t-r)^2} \sqrt{1 + (t + r)^2}}$,
the {\it conformal metric} and its domain of definition are obtained in the form
\[
g = \Omega^{2}\,\hat{g} = - d\tau^2 + d\chi^2 + \sin^2\chi\,d\sigma^2,
\quad \hat{M} = \{\chi \ge 0, \,\, |\tau \pm \chi| < \pi\}.
\]
This metric, the conformal factor, and the underlying manifold smoothly extend to yield
{\it conformally compactified Minkowski space} with manifold
\[
M =  \{\chi \ge 0, \,\, |\tau \pm \chi| \le \pi\} = \hat{M} \cup {\cal J}^{\pm} \cup i^0 \cup i^{\pm}.
\] 
The two components ${\cal J}^{\pm} = \{|\tau \pm \chi| = \pi\}$ of the {\it conformal boundary},
on which $\Omega = 0$, $d\Omega \neq 0$, represent {\it future and past null infinity},
they are generated by the future and past endpoints respectively acquired by  the null geodesics.  They are null hypersurfaces with respect to the {\it conformal metric} $g$. 

The two points $i^{\pm} = \{\tau = \pm \pi, \,\chi = 0\}$, where  $\Omega = 0$, $d\Omega = 0$,
and $Hess_g\Omega = - g$, represent the future and past endpoints of the time-like geodesics and thus  {\it future and past time-like infinity}.  The space-like geodesics run in both directions towards {\it space-like infinity}, represented by the point $i^0 = \{\tau = 0, \,\chi = \pi\}$, where  
$\Omega = 0$, $d\Omega = 0$, and $Hess_g\Omega = g$.
By adding this point, the Cauchy hypersurface $\{t = 0\} = \{\tau = 0, \,\,0 \le \chi < 0\}$ of Minkowski space with the metric induced by $g$ extends to  the sphere $\mathbb{S}^3$ endowed with its standard metric.

\vspace{.3cm}

The process of extending the {\it differential structure} and the {\it conformal structure}
of a given space-time  $(\hat{M}, \hat{g})$ to obtain a smooth (resp. $C^k$ with $k$ sufficiently large) conformal extension  $(M, g, \Omega)$ with boundary $ {\cal J}$ so that 
$M = \hat{M} \cup {\cal J}$, $g = \Omega^2\hat{g}$ on $\hat{M}$ and $\Omega = 0, d\Omega \neq 0$ on ${\cal J}$ as observed above was largely generalized in \cite{penrose:1963}. It was suggested that it applies to many solutions of Einstein's field equations. In the case of solutions which satisfy Einstein's vacuum field equations near ${\cal J}$ it turned out that ${\cal J}$ (consisting then  in general of two components ${\cal J}^{\pm}$) is in fact a null hypersurface for the conformal  metric $g$ that represents (future and past) null infinity. In the situations considered by the authors mentioned above it gives the precise fall-off behaviour required in the asymptotic analysis and it largely simplifies the latter by the possibility to use,
 if ${\cal J}^{\pm}$ is sufficiently smooth,
 local differential geometry instead of taking complicated limits.

In the 1960's , 1970's a large number of articles analyzed the geometrical and physical implications of the new picture  
and various concepts related to ${\cal J}^{\pm}$ were discussed: The Bondi mass, the radiation field, 
 the BMS group, $\ldots$, see e.g. \cite{Geroch: 1977}, \cite{hawking:ellis:1973}
\cite{penrose:1965}, \cite{penrose:1974} and the literature given there.
However, while various concepts seemed to find a natural home in the new picture,
it was not universally accepted by all workers in the field.  
Many 
competing and conflicting aspects are to be considered:

\noindent
$-$ 
questions of mathematical generality, 

\noindent
$-$ the definability and properties of physical concepts,

\noindent
$-$ the capacity to model the various physical situations of interest, 

\noindent
$-$ sharpness of results  and avoidance of redundancies,

\noindent
$-$ existence of solutions to the field equations with the desired asymptotics,

\noindent
$-$ numerical or analytical calculability of  observation related quantities. 

\noindent
Without stating it explicitly in each case, the following discussion will touch, in one way or other, on most of these points.

\section{Asymptotic  smoothness and peeling}

\noindent
R. Penrose \cite{penrose:1965} analyses the behaviour of the conformal Weyl curvature 
$C^{\mu}\,_{\nu \lambda \rho}[g]$ of the conformal extension of an asymptotically simple vacuum solution $(\hat{M}, \hat{g})$ in coordinates $x^{\mu}$ adapted to $g_{\mu \nu}$.  He gives an argument that $C^{\mu}\,_{\nu \lambda \rho}[g]$ vanishes at ${\cal J}^+$, assuming  the conformal extension is sufficiently smooth and ${\cal J}^+$ is diffeomorphic to 
$ \mathbb{R} \times \mathbb{S}^2$. 
Because it involves implicit assumptions on the smoothess of  the conformal extension, it is difficult to assess the precise range of validity of this argument. It certainly works if $(M, g, \Omega)$ is of class $C^4$, but weakening this assumption is a very delicate matter (the argument as reconsidered in  
\cite{friedrich:tueb} starts from smoothness assumptions stronger than those in \cite{penrose:1965}). It will be seen below that the vanishing of $C^{\mu}\,_{\nu \lambda \rho}[g]$
at null infinity  is  in fact necessary for the smoothness of the conformal extension.

The vanishing of the Weyl tensor of the conformal metric at ${\cal J}^{\pm}$ is thus related to the smoothness of the conformal fields. But what does that mean in terms of the physical fields ? It  turns out that it is directly related to the very specific fall-off behaviour of the Weyl tensor  
$C^{\mu}\,_{\nu \lambda \rho}[\hat{g}]$ at null infinity which is known as 
{\it Sachs peeling}. This has been suggested by
R. Sachs \cite{sachs:1961} to be characteristic for the asymptotic behaviour of self-gravitating isolated systems. 

Assume that the conformal extension $(M, g, \Omega)$ with $M = \hat{M} \cup {\cal J}^+$ and $g = \Omega^2\,\hat{g}$ on $\hat{M}$  is of class $C^k$, $k \ge 3$. Let the  function $u$ on 
$\hat{M}$ with $du \neq 0$ define a family of null hypersurfaces $\{u = const.\}$
in $\hat{M}$ that approach ${\cal J}^+$ at space-like surfaces $\sim \mathbb{S}^2$ and 
let $\hat{r}$ be an affine parameter along the future directed null geodesics generating 
the hypersurfaces $\{u = const.\}$ so that  $\hat{r} \rightarrow \infty$ at  ${\cal J}^+$.
Denote by $\{\hat{\kappa}_A\}_{A = 0, 1}$ a spin frame  so that 
$\hat{e}_{AA'} = \hat{\kappa}_A\,\bar{\hat{\kappa}}_{A'}$ is a $\hat{g}$-pseudo-orthonormal frame with   $\hat{e}_{00'} = grad_{\hat{g}}u$ and assume that $<\hat{e}_{00'}, d\hat{r}> \,= 1$.

It can then be shown that the conformal Weyl tensor $C^{\mu}\,_{\nu \lambda \rho}[g]$
(in coordinates  adapted to $g$) vanishes on ${\cal J}^+$ if and only if  the components 
$\hat{\Psi}_k$ of the conformal Weyl spinor corresponding to $\hat{g}$ 
in the spin-frame $\hat{\kappa}_A$ 
have the Sachs peeling property, i.e. they satisfy, {\it with integer powers of 
$\hat{r}$}
\[
\hat{\Psi}_k = \hat{\psi}_k\,\hat{r}^{k - 5} + o(\hat{r}^{k - 5}), 
\quad \quad k = 0, 1, \ldots 4, \quad \quad \mbox{as} \quad \hat{r} \rightarrow \infty,
\]
where the $\hat{\psi}_k$ can be regarded as function of class $C^{k - 3}$ on ${\cal J}^+$
(see \cite{penrose:1974} for details). We note that the function $\hat{\psi}_4$ is interpreted as the 
{\it radiation field on ${\cal J}^+$}.

\vspace{.1cm}
 
Not everybody shared the opinion that asymptotic simplicity encodes the  
fall-off behaviour of self-gravitating isolated systems in an appropriate way.
Workers who studied  equations of motions and tried to calculate the radiation escaping from the system found it difficult to verify Sachs peeling  in their settings. This provoked doubts, 
questions and gave rise to heated discussions (see \cite{ehlers:1979} and the references in
\cite{chrusciel-maccallum-singleton:1995}). In the following years many colleagues who follow the subject only by hearsay still seem to consider peeling as a dubious concept and
D. Christodoulou and S. Klainerman \cite{christodoulou:klainerman:1993} write in 1993:
{\it `$\ldots$  it remains questionable whether there exists any  non-trivial solution of the field equations that satisfies the Penrose requirements. Indeed, his regularity assumptions translate into fall-off conditions of the curvature that may be too stringent  and thus may fail to be satisfied by any solution that would allow gravitational waves'}.

Regardless of its geometric elegance, at the time the concept of asymptotic simplicity was just a proposal based on (profound) guess work. It then appeared natural to demand that
{\it the fall-off behaviour of gravitational fields at null infinity should not be a matter of  guesses but should be derived by achieving  precise control on the evolution process}.
But this leaves the question: {\it Which are appropriate situations from which the fields should  evolve and what precisely is to be achieved} ?
Moreover, far into the 1970's only local in time results had been obtained in the general analysis of the Cauchy problem for Einstein's field equations \cite{choquet80}, \cite{hughes:kato:marsden}.

\vspace{.1cm}

 The first results aiming at the long term evolution of gravitational fields appeared in the early 1980's:

\vspace{.1cm}

Following  Y. Choquet-Bruhat's \cite{foures-bruhat} treatment of the Einstein's vacuum equations as non-linear system of wave equations for the coefficients of the `physical' metric, D. Christodoulou and N. O'Murchadha \cite{christodoulou:o'murchadha:1981}
analysed the  {\it boost problem}.
They  showed that any asymptotically flat initial data for Einstein's vacuum field equations have a development which includes complete space-like surfaces boosted relative to the initial surface. Future or past complete null geodesics, however,  were not under control yet.

\vspace{.1cm}

H. Friedrich \cite{friedrich:1981a}, \cite{friedrich:asymp-sym-hyp:1981b}
studied representations of the Einstein equations in terms of the conformal fields $g$, 
$\Omega$ and derived fields, referred to as {\it conformal Einstein  equations}, and introduced new {\it hyperbolic reductions}, aiming at precise and general  existence results on solutions admitting smooth conformal boundaries.  

\vspace{.1cm}

To avoid permanently switching from one setting to the other,
I shall  in the following  give up chronological orders (which can roughly be reconstructed from the references) and sketch in outline the second  and then the first line of this research and the results obtained.  I shall try to put them into perspective  and consider questions which I think are more relevant than the one in the title.

To keep this article at a reasonable length I clearly have to ignore many contributions, for which I apologize. For the many omitted details of this highly technical subject I refer to the original articles, in particular to \cite{bieri:zipser:2009}, \cite{christodoulou:klainerman:1993}, \cite{lindblad:rodnianski:2004}
with regard to  the first approach and to \cite{friedrich:beyond:2015} and the references given there  in the case of the second one.

\section{The hyperboloidal Cauchy problem}

The first results which show that the smoothness of a conformal boundary can be preserved as a consequence of the field equations was obtained by solving the {\it hyperboloidal Cauchy problem}
\cite{friedrich:1983}.

A space-like hypersurface  $S$ in the conformal extension of an asymptotically simple space-time 
$(\hat{M}, \hat{g})$ is referred to as a {\it hyperboloidal hypersurface} if it extends smoothly to ${\cal J}^+$ and is also space-like there. The set $\partial S = S \cap {\cal J}^+ \sim \mathbb{S}^2$ then defines a boundary of $S$
at which $\Omega = 0$ and $\,d\Omega \neq 0$ (there could be inner boundaries as well but we will not be interested here in those). If $\hat{h}_{\alpha \beta}$ and  
$\hat{\chi}_{\alpha \beta}$ denote the first  and the second  fundamental form induced by $\hat{g}$ on 
$\hat{S} = S \cap \hat{M}$, then, in marked contrast to the behavior of the mean intrinsic  curvature of asymptotically flat standard Cauchy data, which must approach zero at space-like infinity, it holds
$|\hat{h}^{\alpha \beta}\,\hat{\chi}_{\alpha \beta}| \ge c = const. > 0$
near the end at infinity represented by $S \cap {\cal J}^+ $.

Examples of such hypersurfaces  in the conformally compactified Minkowski space considered above are given by the sets $\{\tau = const. \neq 0\}$ which comprise, in particular, the extension of the unit hyperbola 
$\{\eta_{\mu\nu}\,x^{\mu}\,x^{\nu} = -1, x^0 > 0\} \subset \{\tau = \frac{\pi}{2}\}$ that motivated the name. There are, of course, many more general examples. For later use we discuss a particular class of hyperboloidal hypersurfaces  in the Schwarzschild space-time with metric  
\[
\hat{g} = - \left(1 - \frac{2\,m}{r}\right)\,dt^2 + \left(1 - \frac{2\,m}{r}\right)^{-1}\,dr^2
+ r^2\,d\sigma^2,
\] 
where $d\sigma^2$ denotes the standard metric on $\mathbb{S}^2$ and we assume $r > 2\,m \ge 0$.
Since all structures will be spherically symmetric, angular coordinates are suppressed.
In terms of the coordinates 
$w = t - r - 2\,m\,\log(r - 2\,m)$ and $\rho = 1/r$ it follows  that $\Omega^2\,\hat{g} = g$ with 
\[
\Omega = \rho, \quad
g = - (1 - 2\,m\,\rho)\,\rho^2\,dw^2 + 2\,dw\,d\rho + d\sigma^2.
\]
These fields extend smoothly to the set $\{\rho = 0,\,w \in \mathbb{R}\}$ which describes the future conformal boundary 
 ${\cal J}^+$ of the Schwarzschild  solution.
The Cauchy hypersurface $\{t = 0, r > 2\,m\}$ is given in this representation by 
\[
w(\rho) =  - \frac{1}{\rho} - 2\,m\,\log\left(\frac{1}{\rho} - 2\,m\right), \quad 0 < \rho < \frac{1}{2\,m}.
\]
Choosing $\rho_0$ with $0 < \rho_0 < \min (\frac{1}{3\,m}, \frac{1}{1 + 2\,m})$ and replacing $w(\rho)$ by the $C^1$ function 
$w_*(\rho)$ with
$w_*(\rho) = w(\rho)$ for $\rho_0 \le \rho < \frac{1}{2\,m}$ and 
$w_*(\rho) = \frac{\partial w}{\partial \rho}|_{\rho_0}\,(\rho - \rho_0) + w(\rho_0)$
for $0 \le \rho < \rho_0$, 
one obtains a spherically symmetric $C^1$ hypersurface  $S'$ which is hyperboloidal. It intersects ${\cal J}^+$ in the same sphere as the outgoing null hypersurface
$\{ w = - \frac{\partial w}{\partial \rho}|_{\rho_0}\,\rho_0 + w(\rho_0)\}$ and approaches a Minkowskian hyperboloidal hypersurface as $m \rightarrow 0$.
Choosing $\rho_0$ small enough and smoothing $S'$ near 
the sphere $\{w = w(\rho_0), \,\rho = \rho_0\}$ 
while preserving its space-like nature we find:

\vspace{.1cm}

\noindent
 {\it For given $R > 2\,m$ there exist smooth hyperboloidal hypersurfaces in the Schwarz\-schild solution which have intersections with the Cauchy hypersurface $\{t = 0, \,2\,m < r\}$ that comprise the set $\{t = 0, \,2\,m < r \le R\}$ and  approach Minkowskian hyperboloidal hypersurfaces as $m \rightarrow 0$}.

\vspace{.1cm}

If the asymptotically simple space-time 
$(\hat{M}, \hat{g})$ considered above solves Einstein's vacuum field equations, the data 
 $\hat{h}_{\alpha \beta}$ and  $\hat{\chi}_{\alpha \beta}$ induced on the hyperboloidal 
hypersurface $S$ satisfy the vacuum constraints on space-like hypersurfaces and have a specific fall-off behaviour at the boundary $S \cap {\cal J}^+$ which allows them to be conformally transformed and smoothly extended to $S \cap {\cal J}^+$  so as to yield  the 1st and 2nd fundamental form induced by the smooth conformal metric $g$ on $S$. We refer to data with these properties as {\it smooth hyperboloidal Cauchy data}. For the associated
{\it hyperboloidal Cauchy  problem} holds the following:

\vspace{.1cm}

\noindent
H. Friedrich \cite{friedrich:1983}: {\it Smooth hyperboloidal Cauchy data develop into a solution
to the vacuum equations which admits a smooth conformal extension ${\cal J}'^+$ in the future
of $S$ whose null generators have past end points on the boundary of $S$.}

\vspace{.1cm}

There is no `smallness condition' required here and no restriction on the topology of $S$
besides orientability and the existence of a boundary representing the asymptotic end.
The `life time' of the solution depends, of course,  on the nature of the data. In general there may be no conformal gauge in which the null generators of  ${\cal J}'^+$ are future complete
(see R. Geroch, G. T. Horowitz \cite{Geroch-Horowitz:1978} for a notion of completeness of null infinity).

There is an important difference here with the formal expansion type analyses considered in previous studies of asymptotically simple solutions. {\it In that case asymptotic smoothness and thus peeling is put in by 
hand all along ${\cal J}^+$. In the present case it is imposed on the initial slice but 
 is then seen to be preserved along ${\cal J}'^+$ as a consequence of the field equations}.

\subsection{Strong non-linear future stability}

More can be said with further assumptions on the data. 

\vspace{.1cm}

\noindent
Friedrich \cite{friedrich:1986b}: {\it The future development of a smooth Minkowskian hyperboloidal initial data set
$(S^*, \hat{h}^*_{ab}, \hat{\chi}^*_{ab})$ is strongly stable: Any smooth hyperboloidal vacuum initial data set $(S, \hat{h}_{ab}, \hat{\chi}_{ab})$ sufficiently close (in suitable Sobolev norm) to $(S^*, \hat{h}^*_{ab}, \hat{\chi}^*_{ab})$ develops into a solution to Einstein's vacuum equations whose causal geodesics are future complete. Moreover, it admits a smooth conformal extension at future null infinity with conformal boundary 
${\cal J}'^{+}$. The extension can be chosen so that   ${\cal J}'^{+}$ is generated by the  past directed null geodesics 
which emanate from a regular point $i^+$ in the  conformal extension 
and have past end point on  the boundary $\partial S$.}

\vspace{.1cm}

It is a remarkable property of the field equations that they force the null generators to meet, under  
the given assumptions, in a regular point $i^+$ that represents future-time-like infinity.
The result generalizes to the Einstein-Maxwell-Yang-Mills equations \cite{friedrich:1991} and to other Einstein-matter systems with conformally covariant matter fields.

I considered the study of hyperboloidal problems in the beginning as a preparation for the standard Cauchy problem, but it  turned out to be as well suited for the numerical calculation of gravitational radiation fields at null infinity  as the standard Cauchy problem. Peter H\"ubner, who pioneered the numerical studies based on the conformal field equations,
 calculated future complete solutions as the ones considered above, including the set  
 ${\cal J}^{+} \cup \{i^{+}\}$ and the radiation field induced on it \cite{huebner:2001}. 
 For further information on the numerics of hyperboloidal initial value problem we refer to
 \cite{frauendiener:2004} and \cite{rinne:moncrief:2013}.

\subsection{Existence of smooth hyperboloidal data }

L. Andersson, P. Chru\'sciel, H. Friedrich \cite{andersson:chrusciel:friedrich:1992} studied the construction of hyperboloidal data $(S, \hat{h}_{ab}, \hat{\chi}_{ab})$ on a 3-manifold with boundary with second  
fundamental  forms $\hat{\chi}_{\alpha \beta}$ satisfying 
$\hat{\chi}_{\alpha \beta} = \frac{\hat{\chi}}{3}\,\hat{h}_{\alpha \beta}$ on 
$\hat{S} = S \setminus \partial S$,
the analogue of assuming time reflection symmetric data in the standard Cauchy problem. 
 The momentum constraint and the assumed geometry require then  $\hat{\chi} = const. \neq 0$ so that the free datum is given by the conformal class of  the physical 3-metric $\hat{h}_{\alpha \beta}$.

Let $\omega \in C^{\infty}(S)$ be a boundary defining function so that 
$\omega > 0$ on $\hat{S}$ and  $\omega = 0$, $d\omega \neq 0$ on $\partial S$ and let 
$h_{\alpha \beta}$  be a smooth Riemannian metric on $S$. 
The ansatz $ \hat{h}_{\alpha \beta} = \phi^4\,\omega^{-2}\,h_{\alpha \beta}$ 
with an unknown scalar function $\phi$ reduces the Hamiltonian constraint to a singular elliptic problem  for $\phi$:

\vspace{.1cm}

\hspace*{2cm}$R[\phi^4\,\omega^{-2}\,h] = - \frac{2}{3}\,\hat{\chi}^2,
\quad \quad \phi \ge \phi_0 = const. > 0 \quad  \mbox{on} \quad S.$

\vspace{.1cm}

\noindent
{\it There exists a unique solution $\phi$ to this problem. It is smooth on $\hat{S}$ but admits in general only a 
polyhomogeneous expansion at $\partial S$, i.e. an asymptotic expansion 
 in terms of the functions
$\omega^i\,(\log \omega)^j$.  The logarithmic terms vanish and the obtained hyperboloidal data are smooth  if and only if one of the following equivalent conditions is satisfied:

\vspace{.1cm}

\noindent
(i) The trace free part of the second  fundamental form induced by 
$h_{\alpha \beta}$ on $\partial S$ vanishes.

\vspace{.1cm}

\noindent
(ii)  The conformal Weyl tensor $C^{\mu}\,_{\nu \lambda \rho}$ calculated from the data on 
$S$ vanishes on $\partial S$.

\vspace{.1cm}

\noindent
(ii) The `asymptotic shear'  of the null geodesic congruence approaching $\partial S$ which defines the Cauchy horizon of the past Cauchy development  of the solution determined by the data vanishes at $\partial S$.}

\vspace{.3cm}

If the fields $\hat{h}_{ab}$ and $\hat{\chi}_{ab}$ satisfy the constraints, they allow us to calculate the conformal Weyl tensor $C^{\mu}\,_{\nu \lambda \rho}[\hat{g}]$ on $\hat{S}$, where $\hat{g}$ denotes the physical solution metric determined from these data. On $\hat{S}$ it is equal to the Weyl tensor $C^{\mu}\,_{\nu \lambda \rho}[g]$, where $g = \Omega^2\,\hat{g}$ with a suitable conformal factor $\Omega$. If $g$ extended smoothly to the set $\{\Omega = 0\}$ 
the tensor $C^{\mu}\,_{\nu \lambda \rho}[g]$ would also extend smoothly. That it should satisfy in fact $C^{\mu}\,_{\nu \lambda \rho}[\hat{g}] = 
C^{\mu}\,_{\nu \lambda \rho}[g] \rightarrow 0$ at $\partial S$ is a non-trivial condition.

That the asymptotic shear vanishes on the conformal boundary of asymptotically simple vacuum solutions  has been observed already by Penrose \cite{penrose:1965}. That this condition is decisive in the  smoothness discussion for hyperboloidal data is again non-trivial.

Most important is the first condition which is given directly in terms of the free data.
It shows that the latter  only need to satisfy asymptotic condition at the boundary $\partial S$ for the data to evolve into a space-time that admits a smooth conformal boundary in its future.
These conditions are easily satisfied.

\vspace{.1cm}

This result opened the way to the  construction of more general smooth hyperboloidal data.
L. Andersson and P. Chru\'sciel generalized the result in two ways \cite{andersson:chrusciel:1996}:
The second fundamental form was only subject to the requirement
$\hat{\chi} = \hat{\chi}_{ab}\,\hat{h}^{ab} = const. \neq 0$ and free data  
were admitted that have polyhomogeneous expansions at $\partial S$. With these conditions they could show:

\vspace{.1cm}

\noindent
{\it
 The solutions to the constraints again admit polyhomogeneous
 expansion at $\partial S$.
The non-vanishing of the conformal Weyl tensor at $\partial S$ again contributes to the occurrence of logarithmic terms in the solutions to the elliptically reduced constraints.
Conditions on the free data can be given under which the hyperboloidal data  extend smoothly to 
$\partial S$.
}

\vspace{.1cm}

\noindent
While the solutions arising from smooth data admit a smooth conformal extension across null infinity, the much more complicated behaviour near null infinity of solution space-times  arising from  general Andersson-Chr\'usciel hyperboloidal data has not been analysed yet.

\vspace{.1cm}

P. T. Chru\'sciel, M. A. H. MacCallum and  D. B. Singleton \cite{chrusciel-maccallum-singleton:1995} studied general formal  Bondi expansions admitting again asymptotic polyhomogeneous expansions. While some Bondi expansions admitting some logarithmic terms had been discussed earlier (see the references in \cite{chrusciel-maccallum-singleton:1995}) they had not been analysed before in such a systematic way\footnote{Shortly after the version arXiv:1709.07709v1 of this article had appeared, a new result concerning the non-linear stability of Minkowski space was posted by P. Hintz and A. Vasy \cite{hintz:vasy:2017} who consider solutions that are polyhomogeneous at null infinity. How these compare with the formal solutions discussed above  still has to be seen (see also the discussion below).}.

\section{Asymptotically simple vacuum solutions}

C. Cutler and R. Wald  \cite{cutler:wald} managed to construct a parameter dependent family of smooth 
asymptotically flat standard Cauchy data  for the Einstein-Maxwell equations on $\mathbb{R}^3$ that are isometric to  Schwarzschild data  in a neighbourhood of space-like infinity. As we have seen above, the developments in time of such data contain smooth hyperboloidal hypersurfaces which carry smooth hyperboloidal initial data. Since the standard Cauchy data constructed by the authors approach Minkowskian standard Cauchy data  for suitable values of the parameter, the hyperboloidal initial data approach Minkowskian 
hyperboloidal initial data. Invoking the strong stability result discussed above, they were able to conclude:

\vspace{.1cm}

\noindent
{\it There exist non-trivial solutions to the Einstein--Maxwell equations 
whose causal geodesics are complete and which admit smooth conformal extensions
with complete null infinity  ${\cal J}^{\pm}$ and 
regular points $i^{\pm}$ that represent past and future time-like infinity.}

\vspace{.1cm}

This was the first demonstration of  the existence of non-trivial solutions to Einstein's field equations with  smooth and complete asymptotics. At the time the data used here looked rather contrived but ten years later they turned out to be special examples of a much larger class of similar data.

\subsection{The Corvino gluing construction}

 J. Corvino  introduces in \cite{corvino:2000} {\it  a general technique which allows him to deform time reflection symmetric, asymptotically flat  vacuum Cauchy data on $\mathbb{R}^3$ (say) outside a prescribed compact set so that they
become isometric to Schwarzschild data in a neighbourhood of space-like infinity and satisfy the contraints everywhere}.

\vspace{.1cm}

This is a most remarkable result. It gives an unexpected freedom to construct solutions to the constraints which are not accessible by earlier methods \cite{bartnik:isenberg}. 
It also sheds new light on the role of the asymptotic ends at space-like infinity (see the discussion below).

\vspace{.1cm}

P. Chru\'sciel and E. Delay \cite{chrusciel:delay:2003} and  J. Corvino and R. Schoen  \cite{corvino:schoen} generalize this result, showing that 
{\it general asymptotically flat  vacuum Cauchy data can be modified outside prescribed compact sets so as  to become in some neighbourhood of space-like infinity 
isometric  to the Schwarzschild or an other static solution in the time reflection symmetric case and isometric  to Kerr or other stationary solutions in the other cases}.  

\vspace{.1cm}

These data have developments in time that are static or stationary near space-like infinity and thus have smooth conformal asympotics there. Generalizing the construction of the hyperboloidal hypersurfaces for the Schwarzschild solution considered above, we conclude that these results 
also provide means {\it to deform given asymptotically flat data, without changing them on a given compact set, so as to become smooth hyperboloidal at their asymptotic end}. 

\vspace{.1cm}

Instead of using this detour via the evolution in time, P. T. Chru\'s\-ciel and E. Delay \cite{chrusciel:delay:2009} directly use gluing techniques to show {\it the existence of 
a class of non-trivial data which are diffeomorphic to Schwarzschild-anti-de Sitter data outside some compact set and thus provide non-trivial hyperboloidal data}
(for the relation between anti-de Sitter type data and hyperboloidal data see \cite{kannar:1996}).

\vspace{.3cm}

Obviously, Corvino's method was crying for an application along the lines of the  Cutler--Wald idea, but as it stood  his method did not allow him to produce data with arbitrarily small masses. 
  In the following years P. Chru\'sciel and E. Delay  \cite{chrusciel:delay:2002} 
  and  J. Corvino \cite{corvino: 2007} managed, however,  to show the existence of continuous families of smooth, non-trivial standard vacuum Cauchy data  which are exactly static or stationary near space-like infinity and approximate Minkowskian standard Cauchy vacuum data. Evolving these data they thus obtained
{\it families of smooth hyperboloidal data approximating Minkowskian hyperboloidal data}.
Also invoking  the strong stability result above on  Minkowskian hyperboloidal developments, they conclude:

\vspace{.1cm}

\noindent
{\it There exist large classes  of non-trivial solutions to the Einstein vacuum field equations with complete and smooth conformal extension ${\cal J}^{\pm}$
at null infinity and regular points $i^{\pm}$  at past and future time-like infinity.}

\vspace{.1cm}

\noindent
Concerning the `largeness' of the class it should be observed that while the data need to be 
close to Minkowskian data, the deformation techniques discussed above leave the original data unchanged on prescribed compact sets.

\vspace{.1cm}

\noindent
Because the points $i^{\pm}$ for any of these solutions are regular, the vanishing of the radiation field on ${\cal J}^-$ or ${\cal J}^+$ would imply that the solutions were flat \cite{friedrich:1986}. If they have non-vanishing ADM mass, however, they have a non-vanishing conformal Weyl tensor. It follows that they have non-trivial radiation content. {\it Any doubts about the existence of radiative solution with smooth Penrose asymptotics have been put to rest by these results}. 

\vspace{.1cm}

\noindent
Of course, being exactly static or stationary in a neighbourhood of space-like infinity
(any such neighbourhood is of infinite spatial extent)  is a strong assumption on the data and having the conformal boundary $C^{\infty}$ instead of $C^k$, with some $k \ge 4$, is a strong requirement. It should be possible to weaken the assumptions and strengthen the result. 
It will be  seen below that vacuum data  which are {\it asymptotically static or stationary (up to sufficiently high order)}  represent  good candidates for this task.

\section{Space-like infinity touching null infinity}

The hyperboloidal Cauchy problem makes a clear distinction between asymptotically smooth
and non-smooth data and the smooth data develop into solutions that admit smooth conformal extensions at their future null infinity. In the standard Cauchy problem the situation at the asymptotic end at space-like infinity
is not so clear. Compactified Minkowski space is smoothly foliated by the slices
$\{\tau = \tau_* = const.\}$. These are hyperboloidal if $\tau_* \neq 0$ while the slice $\{\tau = 0\}$ 
is  asymptotically Euclidean, extending to the regular point $i^0$ that represents space-like infinity.
When $m_{ADM} > 0$, conformal extensions in which space-like infinity is represented by a regular point do not exist and the transition from an asymptotically Euclidean slice to hyperboloidal slices is in general more complicated. 

To understand possible obstructions to asymptotic smoothness arising in standard Cauchy problems we need to analyse in detail the structure of  solutions in {\it a domain where space-like and null  infinity come close to each other}. In the physical standard representation of the metric, in terms of which the structures referred to are at infinity, it is not clear what should be meant by this and the analysis requires complicated limits.
On the other hand, conformally compactified Minkowski space, in which space-like infinity is represented by the one point $i^0$ is not  a good guide if $m_{ADM} \neq 0$. In such a picture the rich structure discussed below would be compressed into one point and it would be impossible to analyse the field equations.

\vspace{.1cm}

The requirement above acquires a concrete meaning in a setting introduced by
H. Friedrich \cite{friedrich:i-null}, where  
space-like infinity is represented by a cylinder $I = ] -1, 1[ \times \mathbb{S}^2$ which should be thought of as a further piece of boundary of the physical space-time.
It intersects an extended  Cauchy hypersurface in the sphere 
$I^0 = \{0\} \times \mathbb{S}^2$
and touches the sets ${\cal J}^{\pm} = \{\Omega = 0, d\Omega \neq 0\}_{\pm} \sim \mathbb{R} \times \mathbb{S}^2$ at the {\it critical sets}
$I^{\pm} = \{\pm 1\} \times \mathbb{S}^2$. All these sets, which define boundaries and edges of the physical space-time manifold $\hat{M}$, are given at a finite location  in a certain type of coordinate system. 
The setting and the gauge, including the coordinates, a $g$-orthonormal frame, and the conformal factor, are determined, apart from some conditions on the initial slice, by the field equations and the conformal structure of its solutions. The conformal factor and thus the location of the prospective hypersurfaces ${\cal J}^{\pm}$  are known explicitly (see \cite{friedrich:GRG:pune:1998} for illustrations 
and  \cite{friedrich:spin-2} for explicit formulas in the case of Minkowski space).
Because the gauge is based on conformal geodesics, the analysis should generalize to more general initial data than the ones considered below. This requires a careful analysis, however, because the cylinder $I$, which is not a part of the physical manifold, is generated by limits of these curves.

\vspace{.1cm}

No smallness conditions are needed here but to analyse the resulting, somewhat special,  initial boundary value problem, it is convenient (though most likely not necessary) to require the initial data to be {\it asymptotically clean} in the sense that they are smooth and admit expansions in terms of powers 
of a radial coordinate $\hat{r}$ with $\hat{r} \rightarrow \infty$ at space-like infinity. Prescribing `free data'  which satisfy this condition, S. Dain and  H. Friedrich \cite{dain:friedrich} analyse the contraints by standard methods. Besides the solutions of the desired form there are also some with terms 
$\hat{r}^k\,\log \hat{r}$, $k \in \mathbb{Z}$ that are related to a non-vanishing linear 
ADM momentum. As in \cite{christodoulou:klainerman:1993}, these are omitted
in the following discussion. 
{\it In a suitable conformal scaling and in suitably adapted coordinates the boundary $I^0$
of the initial slice at space-like infinity is a sphere at a finite location, 
 and the data for the conformal field equations extend smoothly to $I^0$}.

\vspace{.1cm}

The reduced conformal field equations are in this setting hyperbolic on $\hat{M} \cup I$ and, 
 if the frame admits a continuous extension, also at null infinity ${\cal J}^{\pm}$.  
{\it The hyperbolicity is lost, however,  
at the critical sets $I^{\pm}$}. That a standard Cauchy problem is underlying the construction is 
reflected  by the fact that {\it the boundary $I$ is a total characteristic}: The system of reduced equations reduces on $I$  to a system of interior equations on $I$. 
As a consequence, it allows us to calculate a formal expansion of the space-time along $I$  in terms of a coordinate $\rho \ge 0$ with $\rho = 0$ on $I$ 
by integrating interior equations on $I$. 
The initial data on $I^0$ for this procedure are provided  by the data for the conformal field equations on the initial slice and their derivatives of all orders  at $I^0$ with respect to the radial coordinate $\rho$. The main observations are: 

\vspace{.1cm}

\noindent
{\it Even when the data on the initial slice are smooth near $I^0$, the solutions on $I$ will in general not extend smoothly to $I^{\pm}$ but develop a polyhomogeneous behaviour.}

\vspace{.1cm}

\noindent 
{\it If the setting is linearized at Minkowski space, so that  the equations reduce essentially to the spin-2 equation, 
the polyhomogeneous behaviour at $I^{\pm}$ spreads along the characteristics represented by 
${\cal J}^{\pm}$} \cite{friedrich:spin-2}. 

\vspace{.1cm}

The occurrence of these logarithmic terms  is not a problem of the setting but a consequence 
of the evolution equations and the structure of the data. We cannot expect the situation to be any better in the non-linear case. That it will not be worse has been confirmed recently by the results of 
\cite{hintz:vasy:2017}. The situation can be improved:

\vspace{.1cm}

\noindent
{\it The logarithmic terms do vanish after suitable changes of  the Cauchy data near $I^0$.
In particular, if the data are {\it static or stationary} near space-like infinity the whole setting is as smooth near 
$I \cup I^{\pm} \cup {\cal J}^{\pm}$ as one could wish}
\cite{acena:kroon}, \cite{friedrich:cargese}. 

\vspace{.1cm}

\noindent
These are just the simplest examples. The point here is, however, not so much the staticity or stationarity of the data  in a full neighbourhood of $I^0$. Decisive is instead the detailed structure of the coefficients of the expansion of the data near $I^0$ in terms of the coordinate $\rho$. It holds in fact:

\vspace{.1cm}

\noindent
{\it It suffices that the data be} asymptotically {\it static or stationary at space-like infinity
(at all orders or up to some prescribed order) for 
 the integration on $I$ to be  (at all orders or up to some prescribed order) free of logarithmic terms at $I^{\pm}$.}

\vspace{.1cm}

In the case of time reflection symmetric data  there is 
a certain amount of evidence that  asymptotic staticity  of  the Cauchy data at space-like infinity
is also {\it necessary} for the non-occurrence of  logarithmic terms  at $I^{\pm}$   \cite{friedrich:i-null}, \cite{j.a.v.kroon:2007}. Less is known about necessary conditions for  the smoothness at $I^{\pm}$ in the case of Cauchy data with non-vanishing second fundamental form.
 It can be expected that {\it asympotic stationarity} of the data at space-like infinity is sufficient for the existence of a smooth conformal boundary at null infinity. 

\vspace{.1cm}

\noindent
{\it In cases in which  sufficient smoothness of the boundary at null infinity can be established, the setting above allows us to perform detailed and explicit calculations  which relate the behaviour of the Cauchy data near space-like infinity $I^0$ to the behaviour of the fields on ${\cal J}^+$ near $I^{\pm}$ where the cylinder at space-like infinity meets null infinity \cite{friedrich:kannar1}}.

 \vspace{.2cm}

All the results mentioned so far  on the existence of `general' solutions admitting smooth conformal extensions were obtained by using the {\it conformal field equations} for certain conformal fields 
$\Omega$, $g$, $\ldots$ $W^{\mu}\,_{\nu \lambda \rho}$ derived from the physical metric
$\hat{g}$ and a conformal factor $\Omega$ subject to certain gauge conditions. An important subsystem
of the equations is given by the conformally covariant Bianchi or spin-2  equation which reads in the vacuum case 

\vspace{.1cm}

\hspace*{3.5cm}$\nabla_{\mu}\,W^{\mu}\,_{\nu \lambda \rho} = 0,$

\vspace{.1cm}

\noindent
where $\nabla$ denotes the connection defined by the conformal metric $g$ and  $W^{\mu}\,_{\nu \lambda \rho}$ the rescaled version of the conformal Weyl tensor 
$\hat{C}^{\mu}\,_{\nu \lambda \rho} = C^{\mu}\,_{\nu \lambda \rho}[\hat{g}]$, i.e.

\vspace{.1cm}

\hspace*{3cm}$W^{\mu}\,_{\nu \lambda \rho} = \Omega^{-1}\,\hat{C}^{\mu}\,_{\nu \lambda \rho}
\quad \quad \mbox{on} \quad \hat{M}.$

\vspace{.1cm}

\noindent
This shows that trying to exploit the conformal properties of the Einstein equations in the most direct way has its advantages and its risks. The  conformal field equations lead to complete and sharp  results in situations in which the solutions to be constructed admit a smooth conformal extension. In cases in which peeling does not hold, however, the unknowns develop a singular behaviour as exemplified above by the situation at space-like infinity. This is reflected by the energy estimates for the symmetric hyperbolic equations of first order for $W^{\mu}\,_{\nu \lambda \rho}$ which is implied by the overdetermined system above in the gauges employed: The integrand in the energy estimates is given by components of the rescaled Bel-Robinson tensor, which become singular if peeling fails.

\section{Non-linear stability of Minkowski space}

The results on the global non-linear stability by
D. Christo\-doulou and  S. Klainerman \cite{christodoulou:klainerman:1993},
L. Bieri and  N. Zipser \cite{bieri:zipser:2009}, 
H. Lindblad and I. Rodnianski \cite{lindblad:rodnianski:2004}, 
H. Lindblad \cite{lindblad:2017}, and, most recently, by P. Hintz and A. Vasy \cite{hintz:vasy:2017}
are less detailed as far as the precise asymptotic behaviour is concerned but much more complete than the results referred to above
in that they start from fairly general asymptotically flat standard vacuum Cauchy data on $\mathbb{R}^3$ and control the past and future completeness of all causal geodesics. 
The results are obtained by working  in terms of the physical metric $\hat{g}$.
In \cite{christodoulou:klainerman:1993} and \cite{bieri:zipser:2009} the 
Bianchi equation $\hat{\nabla}_{\mu}\,\hat{C}^{\mu}\,_{\nu \lambda \rho} = 0$
for the conformal Weyl tensor  also plays an important role, its conformal properties are exploited  indirectly and the properties of the Bel-Robinson tensor are used extensively to derive the relevant estimates. The work in \cite{hintz:vasy:2017}, \cite{lindblad:rodnianski:2004} and  \cite{lindblad:2017} is based again on the representation of the Einstein equations as a system of wave equation obtained by imposing a wave gauge. Only some observations of relevance for our discussion will be presented here.

\vspace{.1cm}

All authors assume the  first and second fundamental form $\hat{h}_{ab}$ and $\hat{\chi}_{ab}$  to be smooth, to be close to Minkowskian data in a well-defined  sense, and 
to satisfy certain fall-off conditions near space-like infinity. In the case of
\cite{christodoulou:klainerman:1993} these read
\vspace{.1cm}

\hspace*{.5cm}$\hat{h}_{ab} = \left(1 + 2\,m\,|x|^{-1}\right)\,\delta_{ab} + o_4(|x|^{-3/2}), 
\quad \quad  \hat{\chi}_{ab} = o_3(|x|^{-5/2})$,

\vspace{.1cm}

\noindent
which  implies the vanishing of the linear ADM momentum. This is not the case for 
the generalization given by Bieri in  \cite{bieri:zipser:2009} which requires

\vspace{.1cm}

\hspace{.7cm}$\hat{h}_{ab} = \delta_{ab} + o_3(|x|^{-1/2}), \quad \quad  
\hat{\chi}_{ab} = o_2(|x|^{-3/2})$ \hspace{.5cm} as $|x| \rightarrow \infty$.

\vspace{.1cm}

\noindent
In both cases $x$ denotes a $\mathbb{R}^3$-valued coordinate near the asymptotic end. 
It is shown:

\vspace{.2cm}

\noindent
{\it The  causal geodesics of the maximal globally hyperbolic solutions determined by these data  are complete and their curvature tensors $\hat{C}^{\mu}\,_{\nu \lambda \rho}$ approach zero asymptotically in all directions. }

For our discussion  the rate at which that happens along null geodesics going out to  null infinity is important.
The constructions are based on level surfaces $H_t$ and $C_u$ of a time  function $t$ and 
a retarded time function $u$. The function $\hat{r} = \hat{r}(t, u) > 0 $ is chosen to satisfy
$Vol_{\hat{g}}(H_t \cap C_u) = 4\,\pi\,\hat{r}^2$ on the spherical intersections of the level surfaces.
Along the null geodesics generating the null hypersurfaces $C_u$ it holds then $\hat{r} \rightarrow \infty$ as they run out to future null infinity. Adapting  the notation for the components of the conformal Weyl tensor used in the discussion of Sachs peeling it follows
\[
|\hat{\Psi}_k| = O(\hat{r}^{-7/2}), \,\,k = 0, 1, \quad  \,\, \,\, |\hat{\Psi}_k| = O(\hat{r}^{k - 5}), \,\,k = 2, 3, 4,\,
\]
in the case of \cite{christodoulou:klainerman:1993} 
and in the case of the generalization given by Bieri in  \cite{bieri:zipser:2009} 
\[
|\hat{\Psi}_k| = o(\hat{r}^{-5/2}), \,\,k = 0, 1, 2, \quad  |\hat{\Psi}_k| = O(\hat{r}^{k - 5}), \,\,k = 3, 4,
\quad 
\]
as $\hat{r} \rightarrow \infty$ along a fixed null generator of $C_u$.
For $k = 3, 4$ the behaviour is thus similar to Sachs peeling while it deviates from it, 
{\it in the case of solutions for which these estimates are sharp},  for $k = 0, 1$. 
What is the origin of these deviations ? 

The result by P. Hintz and A. Vasy \cite{hintz:vasy:2017} is based on Cauchy data 
that admit a  polyhomogeneous expansion at space-like infinity
and it is shown that 
the solutions are polyhomogeneous at null infinity. 
The logarithmic  terms observed  at the critical set  in the analysis of space-like infinity outlined above should contribute to them and the logarithmic term observed 
on the initial slice in the construction of clean initial data \cite{dain:friedrich} should contribute to the stronger deviation in the case of  \cite{bieri:zipser:2009}. It would be interesting to know which of the logarithmic terms observed in \cite{andersson:chrusciel:1996},  
\cite{chrusciel-maccallum-singleton:1995}, \cite{hintz:vasy:2017}  can in fact occur in the solutions considered in \cite{christodoulou:klainerman:1993} and  \cite{bieri:zipser:2009}. 
Moreover,  if the data are specified in terms of weighted Sobolev spaces or if they are just required to be smooth and to satisfy the fall-off at space-like infinity indicated above, there is a large freedom to have besides logarithmic terms all kinds of other terms in the data
which may spoil the smoothness of any conformal extension at null infinity at higher orders. 

The question whether the coefficients coming with the logarithmic terms 
or any other non-smoothness properties admitted by the Sobolev norms
are of any physical significance or just represent `noise' is left untouched.

\section{Approximative solutions}

Analytical and numerical approximations are of utmost importance, because they allow us to relate 
measured data to theoretical results. Nevertheless, I shall only make some sketchy remarks about certain aspects related to my topic.

The analytical approximation theory designed to produce quantitative results on the radiation generated e.g. by the merger of black holes should have a counterpart formulated  in terms of the Cauchy problem.  
 It certainly would be most useful if more were known about this.
The relation between the abstract and the approximative analytical understanding  is not easily extracted from the literature, however,  because the latter usually  immediately intertwines
 general considerations with the technical details of the approximation method. 
 
L. Blanchet and  T. Damour  note in their extensive work on  approximation methods begun in \cite{blanchet:damour:1986} a difficulty to verify the peeling behaviour (see also the remarks by D. Christodoulou \cite{christodoulou:2002}). On the other hand they impose near space-like infinity conditions to exclude radiation coming in from the infinite past. As stated more explicitly by  L. Blanchet  \cite{blanchet:1987} and  T. Damour and B. Schmidt \cite{damour:schmidt:1990}, it amounts 
to requiring the solutions to be stationary near space-like infinity. In these articles are also given arguments that the solution will then admit a smooth conformal extension near space-like infinity (a gap in the  argument in \cite{damour:schmidt:1990} has been filled in by S. Dain \cite{dain:stationary}). 

This raises the question why the asymptotic  smoothness should  be lost at a later (in 
retarded time) stage of the development. Is this due to the eruptive behaviour of some matter system ? 
But why should it happen  in the pure vacuum case, e.g. in the merger of black holes ?
 Corvino's result and its generalizations were not available at the time. But even if they had been known
 already, it is hardly conceivable that something like  the gluing procedure could be realized in the context of an approximation method. Is the loss of asymptotic smoothness possibly  just  
an artifact of the approximation method ?

\vspace{.2cm}

To keep their numerical grids finite, most relativists who develop numerical $3 + 1$ codes for the standard Cauchy problem to calculate binary black hole merger (say)  wave forms use cut-off procedures and essentially ignore space-like and null  infinity, thus also the finer details of the  asymptotic behaviour there. The radiation field, originally  defined at null infinity, is calculated  only approximately 
at a finite, somewhat arbitrary  location (see, however, the work by F. Beyer et al
\cite{beyer:doulis:frauendiener:whale:2012} and J. Frauendiener and J. Hennig \cite{frauendiener:hennig:2017} which takes first steps towards calculating entire solutions determined by asymptotically flat Cauchy data). This cut-off deletes a neighbourhood of  ${\cal J}^{\pm} \cup i^0$ which is of infinite extent as measured in terms of affine parameters on the outgoing null geodesics. Nevertheless, for the time being the results seem to be satisfactory.

Characteristic  or  hyperboloidal  Cauchy problems
with data prescribed on null or space-like hypersurfaces that extend to null infinity have also been solved numerically. The freedom  in the choice of data near 
${\cal J}^+$ can be used  to extend the data smoothly to ${\cal J}^+$. The difference with the standard Cauchy problem is that the wave form extraction can be done at the well and uniquely defined hypersurface ${\cal J}^+$.

We note that in all three cases, the standard Cauchy problems, the hyperboloidal problems, and in characteristic initial value problems, there is a large arbitrariness in choosing the data near the asymptotic end of the initial slice.

\section{A different type of  approximation}

The following two results are of particular interest in our discussion.
P. Allen and I. Stavrov Allen \cite{allen:stavrov/allen:2017}  show:

\vspace{.1cm}

\noindent
 {\it Polyhomogeneous, asymptotically hyperbolic, constant mean curvature 
data of Andersson-Chru\'sciel type for the vacuum Einstein equations can be approximated arbitrarily closely in certain H\"older norms by smooth hyperboloidal constant mean curvature vacuum data.}

\vspace{.1cm}

\noindent
J. Corvino and R. Schoen \cite{corvino:schoen}  state another density result by which:

\vspace{.1cm}

\noindent
{\it Asymptotically flat initial data for the vacuum Einstein equations on a three-manifold 
$\hat{S}$ 
can be approximated by data on $\hat{S}$ which agree with the original data inside a given compact domain, and are  in a given end identical to that of a suitable Kerr slice (or identical to a member of some other admissible family of solutions) outside a large ball. }

\vspace{.1cm}

\noindent
It should be noted that these approximations are controlled in terms of Sobolev norms which are weighted so that 
$\hat{h}_{ab} - \delta_{ab}$ and $\hat{\chi}_{ab}$ and the corresponding approximating data are consistent  with the fall-off behaviour required for these fields in \cite{christodoulou:klainerman:1993}, for example. We state the results here without further details, since different  function spaces may be needed if one wants to answer the most interesting  question provoked by these results:

\vspace{.1cm}

\noindent
{\it Are the asymptotically simple vacuum solutions  in some sense dense in a set of asymptotically flat vacuum solutions as considered in the 
non-linear stability results above ?}

\vspace{.1cm}

If a definite answer could be given to this question
the `mathematical exercises' discussed in the previous sections could be brought to a conclusion. 
The proof of any such density results should completely clarify the situation.
`Peeling or not peeling'  may just become a matter of deciding between technically more or less convenient representations. It may also show that the questions raised in our discussion of analytic and numerical approximations may be essentially harmless.
The situation would be somewhat reminiscent of the introduction of 
$L^2$-Hilbert spaces in the analysis of hyperbolic equations 
as an intermediary  step towards obtaining existence results about smooth solutions.
A positive result should provide interesting information about the precise way in which solutions that are `rough' at null infinity are approached by solutions that have smooth ${\cal J}^+$ and in which way concepts that are easily defined on smooth ${\cal J}^+$'s can be transferred (if at all) to concepts on 
the `rough' future null infinity. 

\vspace{.1cm}

On the other hand, showing that such a density property cannot hold  should explain in which sense rough asymptotics can be superior to smooth asymptotics. It should give information about physical systems of interest which cannot be modeled in the class of asymptotically simple solutions and tell us precisely  what  is lost if we restrict to asymptotically simple solutions. It should further give answers to the following questions.

\vspace{.1cm}

 The logarithmic terms at null infinity mentioned so far 
 come with certain coefficients.  Which information is encoded in these coefficients ? 
What is the physical information in the coefficient in the 'free data' underlying the construction of hyperboloidal or Cauchy data which must be set to zero to get rid of logarithmic terms ?
Doing so, does it lead to a loss of essential  physical information ? 
What is the role of the logarithmic term on the initial slice which is related to the linear ADM momentum ? 
What is its effect on the structure of the radiation field or other quantities of physical interest on null infinity ?
If it can be shown that the logarithmic terms found at the critical sets are indeed related to the deviation of the data from being asymptotically stationary, could this be interpreted as saying that radiation coming in from past null infinity has to be excluded (up to some order) close to space-like infinity to achieve asymptotic smoothness (of a prescribed  order) ?

\section{Isolated systems as part of our cosmos}

Most  of the considerations above are related to the structure of the 
{\it asymptotic end at space-like  infinity} in the standard and to the 
{\it asymptotic  end at null  infinity} in  the hyperboloidal Cauchy problem.
But in the {\it `real world'} of our cosmos a system which we would like to see as one representing a {\it self-gravitating isolated system}  does not possess anything like an asymptotic end at space-like or null infinity.

\vspace{.1cm}

\noindent
The best we can do is to consider an open, relative compact subset $S'$ of a time-slice $S$ of our cosmos with the Cauchy data $d'$ induced on it  so that its domain of dependence $D(S')$ contains the essential part of the object and any related process of interest but no further comparable objects. As a next step we could try  to  {\it attach an asymptotically flat or hyperboloidal end} smoothly to 
$(S', d')${\footnote{G.F.R. Ellis  \cite{ellis:1984}  suggested  to abandon the asymptotically flat model  
and to  introduce instead  in an ad hoc fashion a spatially compact time-like hypersurface ${\cal T}$ 
to cut off `the system of interest' from the ambient universe. We refer the reader to \cite{friedrich:cargese} for a  discussion of the difficulties of this idea and to \cite{friedrich:ibvp-unique}
for the unresolved (possibly unresolvable) difficulties with the underlying initial boundary value problem for Einstein's field equations.}.

\vspace{.1cm}

\noindent
If $S$ is assumed to be compact and all the matter fields are ignored,
space-time engineering as suggested by P.T. Chru\'sciel. J. Isenberg, D. Pollack \cite{chrusciel:isenberg:pollack} allows us in fact to glue an asymptotically flat or hyperboloidal end to $S$. The resulting standard Cauchy  data  will contain, however,  a huge number  of other systems, which we wish to exclude, and there may even be something like a minimal surface close to the location of the gluing process.  What we want is closer to the results of S. Czimek \cite{czimek:2016}, who constructs asymptotically flat extensions of solutions to the vacuum constraints on a compact manifold with boundary that have vanishing mean extrinsic curvature. Since this is done, so far, only for data close to Minkowskian data, the result, as it stands, does not cover  systems containing e.g.  black holes. Another possibility might be to extend the  set $(S', d')$, possible after some modification close to its boundary, along the lines of R. Bartnik's  parabolic constructions of constrained data discussed in  \cite{bartnik:isenberg}.
All these studies suggest that it is not too far-fetched to assume, as we shall do,  that $(S', d')$ can be  embedded  isometrically into some asymptotically flat standard Cauchy set. 
Our earlier discussions then show that it can equally well be embedded into smooth hyperboloidal initial data sets.

\vspace{.1cm}

\noindent
Whatever one does, while suggested by a large class of static or stationary exact solutions and while being consistent with the field equations under much more general assumptions,
{\it the asymptotically flat or hyperboloidal end is a just  figment}. One of its main virtues is to 
allow the field equations themselves  to construct  in  a marvelously effective way  a null infinity and the radiation signal  to unfold while approaching that null infinity along outgoing null rays.

As we have seen, however,  there exists a huge freedom to choose or modify  asymptotic  ends of initial data near space-like or null  infinity, leaving the data unchanged on a large interior subset. The transition from  one choice to another one will affect the solution space-time in a neighbourhood of null infinity and the gluing region may give rise to
some {\it spurious radiation} 
which we may not accept  as being  associated with the system we wish to study (an phenomenon  well known to numerical relativists). But who says that some such radiation had not been encoded already in the first choice of data ?
Those of us who spent much of their time working on  the Kerr family or other  real analytic  stationary solutions may find it strange,  but if we consider ends of class $C^{\infty}$ or $C^k$ 
we have to face the fact that {\it there simply does  not exist  `the' correct asymptotic end}.
 We can only hope to optimize the situation in some sense.

\vspace{.1cm}

This raises the question:  {\it How to make the best use of the freedom to choose 
the  asymptotically Euclidean or hyperboloidal end ?}
 And more specifically: {\it To what extent do radiation fields and other physically relevant quantities defined at null infinity  depend on the precise  structure of the initial data near space-like (or null) infinity ?}  

\vspace{.1cm}
 
 Analysing these questions should give deeper  insight into some physics because the answers 
will depend very much on the nature of the system we wish to  model and the conclusions we want to draw.
The inspiral, merger, and ring down of black holes can be expected to be accompanied by
{\it strong,  well-pronounced signals in the domain of dependence of some interior domain}. 
Reasonable changes near the ends at space-like or null  infinity are likely to have little effect on these
and we may well choose the end to be stationary.
In scattering problems involving weak fields, however,  we can, in principle,  only toy with the data on ${\cal J}^-  \cup \, i^-$ (see \cite{christodoulou:2009} and, for a full treatment of a neighbourhood of $i^-$, also  \cite{chrusciel-paetz-2013}, \cite{friedrich:taylor-at-i}). 
Then we have to watch how things develop  
at  space-like infinity and what comes out at  ${\cal J}^+$. There are data on ${\cal J}^-$ which make the end at  space-like infinity stationary but whether they will be useful in this context depends very much on the questions to be answered.

\vspace{.5cm}

\noindent
Acknowledgements: I would like to thank L. Bieri, J. Corvino, P. Chru\'sciel, D. Garfinkle, 
J. Isenberg, R. Schoen, and R. Wald for discussions.
Part of this work was supported by the program on Geometry and Relativity  at the Erwin Schroedinger Institute, Vienna. I wish to thank the Institute  for support, hospitality and the opportunity to talk about the subject with various  colleagues.

}

\end{document}